\documentclass[a4paper,100pt]{article}

\usepackage[utf8]{inputenc}
\title{A More Realistic Z-pinch Snowplow Model}
\author{Miguel Cárdenas\\Universidad de Playa Ancha\\Av. Playa Ancha 850, Valparaíso, Chile\\e-mail: miguel.cardenas@upla.cl}

\begin{document}
\date{}

\large
\maketitle
\begin{abstract}
We introduce an extended snowplow model for Z-pinch experiments that accounts for partial particle entrainment and current loss during contraction. We applied the methods to a specific case.
\end{abstract} 

\section{Introduction}
The macroscopic physical behavior of the plasma in Z-pinch experiments is governed by several key control parameters. Assuming infinite plasma conductivity, eight control parameters summarize the macoscopic physical behavior of the plasma. Dimensional analysis shows that the eight physical parameters can be arranged into three independent dimensionless $\Pi$-terms, capturing the core macroscopic behavior of the experiment \cite{bridgman}\cite{langhaar}\cite{barenblatt}\cite{gukhman}\cite{cardenas1}.
 
Even with this simplification, the need to mathematically model these experiments remains. The computationally most economical scheme that, nevertheless, provides an estimation of the plasma temperature is the snowplow model \cite{rosenbluth}. This model can be formulated as two coupled, non-linear integro-differential equations that depend on two dimensionless $\Pi$-terms. Numerical solution of the equations yields the third dimensionless $\Pi$-term. Practically, this system is self-contained and easy to simulate
 \cite{cardenas1}\cite{rosenbluth}\cite{cardenas2}\cite{cardenas3}.
Despite this, the snowplow model still presents a weakness. It is based on the assumptions of complete plasma particle entrainment and no current loss. But these hypotheses are unlikely to be validated in practice. 

This study improves upon previous work by introducing a more realistic model that avoids these improbable assumptions. We now assume that only a fraction of the plasma particles participates in the contraction process and that the current sheath conducts only a portion of the total source current. Implementing these ideas results in a mathematical structure formally identical to the original snowplow equations. However, the dimensionless $\Pi$-terms in the snowplow equation are now unknown. We must rely on experimental measurements to determine them. Consequently, the price of extending the snowplow model is that it is no longer self-contained and requires experimental feedback. As a counterpoint, the upgraded model provides a plausible explanation for the elevated plasma temperatures observed in several Z-pinch experiments. A thought-provoking example on this topic is included at the end of this article. 

The paper is organized as follows: In Section 2, we outline the extended model's key components and introduce the relevant equations. In Section 3, we discuss our results and illustrate the methods with a specific example. In Section 4, we highlight key aspects.  

\section{The Model and Its Equations}

Assuming infinite conductivity of the working gas, the snowplow model offers the most straightforward and fundamental description of plasma dynamics in Z-pinch devices. This assumes the magnetic piston collects all encountered particles as it moves towards the Z axis. The particles then collect in a thin current-carrying layer on the moving piston \cite{rosenbluth}.

In practical experiments, actual plasma dynamics will likely deviate from this theoretical model. Consequently, amending the original postulates of the snowplow model can enable a better approach to the dynamics of real-world experiments. For instance, we can assume the contraction process affects only a fraction $c_1$ of the total number of atoms $N_0$ within the cylinder, while the remaining $(1-c_1)$ fraction is left behind. Furthermore, the working gas conducts a fraction $c_2$ of the total source current $I$; the remaining $(1-c_2)$ constitutes leakage loss. 

This is easy to integrate into the snowplow model \cite{cardenas2}\cite{cardenas3}. Through algebraic manipulation, we derive the following upgraded snowplow equations for the radius of the current sheath $r(t)$ and the total source current $I(t)$:

\begin{equation}
\left(\frac{d^2r}{dt^2}\right)=\frac{6\,r^2\, (dr/dt)^2-3\,(\alpha^\prime)^2 \,I^2-4\int_0^t (1/r)\,(dr/dt')^3 \,dt'}{3\,r\,(1-r^2)},
\end{equation}
\\

\begin{equation}
\left(\frac{dI}{dt}\right)=\frac{1-\int_0^tIdt'+\beta^\prime\ (I/r)\,(dr/dt)}{1-\beta^\prime\ln{(r)}}
\end{equation}
where

\begin{equation}
(\alpha^\prime)^2=\left(\frac{c_2^2}{c_1}\right)\alpha^2,
\end{equation}

\begin{equation}
\beta^\prime=c_2\,\beta
\end{equation}
with the dimensionless groups $\alpha^2$ and $\beta$ defined as

\begin{equation}
\alpha^2=\left(\frac{\mu_0 C_0^2  V_0^2}{4\pi^2  r_0^4\rho_0 }\right),
\end{equation}

\begin{equation}
\beta=\left(\frac{\mu_0 l_0}{2\pi L_0}\right).
\end{equation}
The physical parameters that define these dimensionless groups are listed below
\begin{itemize}
\item The vacuum permeability, $\mu_0$.
\item The inner radius of the cylindrical container, $r_0$.
\item The length of the cylinder, $l_0$.
\item The cylinder filling gas density, $\rho_0$. 
\item The total parasitic inductance of the setup, $L_0$.
\item The total capacity of the bank of capacitors, $C_0$.
\item The initial voltage across the bank of capacitors, $V_0$.
\end{itemize}
The snowplow equations, eqs. (1) and (2), have already been written in dimensionless form; time $t$ is given in units of 
$\sqrt{L_0C_0}$, the radius of the current sheath $r$ is given in units of $r_0$ and the total source current $I$ is given in units of 
$\sqrt{C_0/L_0}\,V_0$. 

Accurately solving the snowplow equations requires determining the constants $c_1$ and $c_2$. Regrettably, this implies a direct, potentially cumbersome, experimental measurement. We will show how to avoid this problem later.

Moving forward with our strategy, the internal energy of the ions $U$ at time $t$ is computed using the following equation:

\begin{equation}
U=E_0\times \left\{c_2\,\frac{\beta^\prime}{(\alpha^\prime)^2}\,\int_0^t\left(-\frac{1}{r}\left(\frac{dr}{dt'}.\right)^3\right)dt'\right\}
\end{equation}
\\
The source energy 

\begin{equation}
E_0=\frac{1}{2}C_0V_0^2
\end{equation}
\\
mantains the correct physical units in eq. (7), since the accompanying factor is dimensionless. The corresponding ion temperature $k_BT$, also expressed in energy units, is given by

\begin{equation}
k_BT=\frac{2}{3}\left(\frac{U}{c_1N_0}\right)
\end{equation}
where the total number of atoms, $N_0$, of the working gas inside the cylinder is calculated as

\begin{equation}
N_0=\frac{\rho_0\,\pi\, r_0^2 \,l_0}{m}
\end{equation}
with $m$ representing the atomic mass of the working gas.

Substituting eq. (7) into eq. (9), we obtain

\begin{equation}
k_BT=\frac{2}{3}\left(\frac{E_0}{N_0}\right)\times \left\{\left(\frac{c_2}{c_1}\right)\frac{\beta^\prime}{(\alpha^\prime)^2}\int_0^t\left(-\frac{1}{r}\left(\frac{dr}{dt'}\right)^3\right)dt'\right\}
\end{equation}
\\
that under a further simplification converts into

\begin{equation}
k_BT=\frac{2}{3}\left(\frac{E_0}{N_0}\right)\times\left\{\frac{\beta}{\alpha^2}\int_0^t\left(-\frac{1}{r}\left(\frac{dr}{dt'}\right)^3\right)dt'\right\}.
\end{equation}

\section{Results and Discussion}
The ion temperature from eq. (12) can, in principle, be computed by numerically solving eqs. (1) and (2) and subsequently substituting the obtained $r(t)$ into the right-hand side of eq. (12). However, this seemingly routine procedure faces a problem. While we know the values of $\alpha^2$ and $\beta$, we do not know the values of $(\alpha^\prime)^2$ and $\beta^\prime$. Determining them requires knowing $c_1$ and $c_2$ in advance. 
Since we do not have enough information to set up and solve eqs. (1) and (2), we cannot obtain their solution either.

Thus, to calculate the ion temperature via eq. (12), we have to use the experimentally measured $r(t)$ curve. Alternatively, fitting eqs. (1) and (2) to the experimental curve yields the coefficients $c_1$ and $c_2$, allowing us to determine $(\alpha^\prime)^2$ and $\beta^\prime$. Having established this,  eqs. (1) and (2) can now be used to proceed with the calculations. Below we present a specific example that illustrates this procedure. 

Let us consider an experimental setup where $E_0 \approx 850\, [J] $, $N_0 \approx 1.16\times10^{20} $ whereas the dimensionless groups formed by the control parameters take the values $\alpha^2=2.56$ and $\beta=2.23$. Additionally, the experiment produced the curve $r_{exp}(t)$. 

Substituting this curve into the right-hand side of eq. (12) yields a temperature $(k_BT)_{exp}=80\,eV$. Furthermore, fitting eqs. (1) and (2) to $r_{exp}(t)$ yields $(\alpha^\prime)^2=2.3$ and $\beta^\prime=0.67$. These values, in turn, result in the coefficients $c_1=0.1$ and $c_2=0.3$.

It is important to highlight that, under optimal conditions with complete particle entrainment and no current losses, \textit{i.e.}
 $c_1=1$ and $c_2=1$, the snowplow model yields a temperature $k_BT=10\,eV$ only.

\section{Closing Remarks}

\begin{enumerate}
\item The macroscopic, measurable physical characteristics of Z-pinch experiments are enterely determined by a small set of foundational control parameters.
\item The modeling process necessitates a preliminary hypothesis regarding the system's operational behavior.
\item We assume the snowplow model, augmented by unknown coefficients $c_1$ and $c_2$. Consequently, parallel experiments are necessary to determine $c_1$ and $c_2$ and solve the model equations.
\item To avoid directly measuring coefficients $c_1$ and $c_2$, we can determine them by measuring $r_{exp}(t)$.
\item Additionally, the curve $r_{exp}(t)$ can be reconstructed using  eqs. (1) and (2).
\item From this point on, estimating the plasma temperature is straightforward.
\end{enumerate}

\end{document}